\documentclass[sts,preprint]{imsart}

\RequirePackage[OT1]{fontenc}
\usepackage{amsthm,amsmath,natbib}
\RequirePackage[colorlinks,citecolor=blue,urlcolor=blue]{hyperref}

\startlocaldefs
\numberwithin{equation}{section}
\theoremstyle{plain}

\endlocaldefs

\usepackage{amsmath,amsfonts,amssymb}

\renewcommand{\le}{\leqslant}

\newcommand{\bsx}{\boldsymbol{x}}
\newcommand{\bsX}{{\boldsymbol{X}}}
\newcommand{\rd}{\,\mathrm{d}}
\newcommand{\mrd}{\mathrm{d}}
\newcommand{\dustd}{\mathsf{U}}

\newcommand{\mc}{\mathrm{MC}}
\newcommand{\qmc}{\mathrm{QMC}}
\newcommand{\mcmc}{\mathrm{MCMC}}

\newcommand{\pn}{\mathrm{PN}}
\newcommand{\prt}{\mathrm{Pr}} 

\newcommand{\hk}{{\mathrm{HK}}}

\begin{document}
\begin{frontmatter}
\title{Unreasonable effectiveness of Monte Carlo\thanksref{t1}}
\runtitle{Comment: yes, there's a role}
\thankstext{t1}{Supported by NSF under DMS-1521145 and  IIS-1837931.
The ideas here also developed in part at SAMSI quasi-Monte Carlo 
meetings, supported by DMS-1127914.}

\begin{aug}
\author{\fnms{Art B.} \snm{Owen}\thanksref{t1}\ead[label=e1]{owen@stanford.edu}}
\runauthor{A. B. Owen}
\affiliation{Stanford University}

\address{Department of Statistics, Sequoia Hall
\printead{e1}.}

\end{aug}

\begin{abstract}
There is a role for statistical computation in numerical
integration.  However, the competition from incumbent methods
looks to be stiffer for this problem than for some of the
newer problems being handled by probabilistic numerics.
One of the challenges is the unreasonable effectiveness
of the central limit theorem. Another is the unreasonable
effectiveness of pseudorandom number generators.  A third
is the common $O(n^3)$ cost of methods based on Gaussian
processes. Despite these advantages, the classical methods
are weak in places where probabilistic methods could bring
an improvement.
\end{abstract}


\end{frontmatter}

\section{Introduction}

I think that the answer to the question in the
authors' title is `yes', despite some challenges that
I will describe.  The title of an earlier version
at arXiv asked about a `role for statisticians 
in numerical analysis'.  There the answer is a resounding `yes'.
That role for statisticians includes developing
Bayesian and frequentist methods, applying them to problems such as integration
and approximation, and then using them to get both point estimates and 
uncertainty quantifications (UQ), such as interval estimates.
Statistical ideas for numerical methods have a long history
and there are exciting new developments too. 
Two examples from \cite{cock:oate:sull:giro:2017:tr} are:
using Bayesian methods to study multiple solutions to
Painlev\'e PDEs, and using those methods to study
an entire computational pipeline taking account of
the fact that some steps are cheap to change, some expensive
and others completely frozen.
Those problems are fascinating and important, underserved
by frequentist methods,  and I expect 
to see good progress on them from Bayesian methods
in the coming years.

The paper focuses on the use of Bayesian methods
to estimate integrals and especially to quantify the uncertainty
in those estimates of integrals.  
This looks like tougher going because the incumbent
methods have some `unreasonable effectiveness' properties
that will be hard to match.
After describing those strengths, I will conclude
by describing areas where the classical methods are weak
providing an opportunity for probabilistic numerics (PN).

First, a few QMC-related remarks: 
The finite order weights in Section 5.4.2 build in an assumption
that the integrand has no interactions whatsoever of
order 3 and up (not just that they are small).  This is considered
quite risky \citep{sloa:2007}. Effective dimension is not
usually defined as a sum of $\gamma_u$. That sum
might not be smaller than $d$.  For a brief history of effective
dimension in QMC, going back to the 1950s, see \cite{effdimpreso}.
The error in higher order digital nets can be reduced
by a factor of about $n^{-1/2}$ by scrambling
the digits. See \cite{dick:2011} for conditions.

The authors have not seen BMCMC used.
Something like that is in the forthcoming paper of \cite{lavi:hodg:2018}.
They use unequal weights 
designed for autocorrelations of the form $\rho^{|i-i'|}$
between observations $f(\bsx_i)$ and $f(\bsx_{i'})$.
As a result they estimate population means
by an unequally weighted sum of sample values.
Their weights correspond to BMCMC if there is a
first order autoregressive posterior distribution.

\section{Inferential basis}

The numerical approaches to integration that we compare
begin by writing the integrand as an expectation of
a quantity $f(\bsx)$ where $\bsx$ has a probability density $p$.  
The integral estimates then take the form 
$$\hat\mu = \sum_{i=1}^n w_if(\bsx_i)$$
where $\bsx_i$ are representative of $p$
in a sense that depends on the method being used.
Weights $w_i^\mc$, $w_i^\qmc$, $w_i^\mcmc$ and $w_i^\pn$ generate 
estimates $\hat\mu^\mc$, $\hat\mu^\qmc$, $\hat\mu^\mcmc$ and $\hat\mu^\pn$,
for Monte Carlo, quasi-Monte Carlo, Markov chain Monte Carlo and 
probabilistic numerics, respectively.
For QMC we ordinarily use methods such as those in \cite{devr:1986} to
make desired non-uniform random variables from uniform ones.  That is
we arrange for $p=\dustd[0,1]^d$, along with any necessary compensating changes to $f$.

For MC, the law of large numbers (LLN) treating the $\bsx_i$ as genuinely
random, gives $\hat\mu^\mc\to\mu$ with probability one. 
For MCMC we also use an LLN  but need additional
assumptions about how the $\bsx_i$ approach their target distribution
and how they mix.  In QMC, the $\bsx_i$ are ordinarily determinstic
points in $[0,1]^d$.  The counterpart to the LLN is that if $f$ is Riemann 
integrable and the star discrepancy
$D_n^*$ \citep{nied:1992} between $\dustd\{x_1,\dots,x_n\}$ and 
$\dustd[0,1]^d$ vanishes then $\hat\mu^\qmc\to\mu$
\citep{nied:1978}.  For PN,  the present paper proves convergence
with probability $1$ under a Gaussian process (GP) model for $f$.

For QMC, the $w_i$ are usually $1/n$.  In some MC methods, $w_i$
is a function of $x_i$. For PN and some other MC methods
each $w_i$ can depend on all of $\bsX = (\bsx_1,\dots,\bsx_n)$.
MCMC usually uses equal weights, often skipping the
first few observations and/or thinning to every $k$'th observation.
In any of these cases we get $\hat\mu=\hat\mu(f,\bsX)$,
a function of both $f$ and $\bsX$.
We then make an error of size $\Delta=|\hat\mu-\mu|$ and
we would like some idea of how large that is.

What does it mean to know $\Delta$?
\cite{diac:1988} begins with a closed form expression for a function,  
and then asks ``What does it mean to `know' a function?''. He then  
discusses  Bayesian numerical analysis, cites some historical references  
and shows how Bayes can recover some well known methods as special cases.  
His question applies with equal force to the error $\Delta=|\hat\mu-\mu|$. 
In what sense is it known (or unknown) when there is a 
precise mathematical expression for it?

For MC and MCMC one usually models $\bsX$ as random
to get a distribution on $\Delta$. For PN, one models $f$ as random for fixed $\bsX$.
It seems compelling from a Bayesian point of view to condition on the observed value  of $\bsX$, 
thereby treating them as known and not random.
The same argument can be made for $f$.  We might view $f$ as a set of bytes describing
a computation or more usefully as some (usually) smooth function describing a quantity
of scientific interest.  When computing $\hat\mu$ however, one such $f$ has been chosen
and even if it had been chosen at random, we could reasonably condition on it.

If we condition on both $f$ and $\bsX$ then $\hat\mu-\mu$ is not
random and it is hard to motivate other values it could have taken
in order to fill up a confidence interval. 
One approach is to treat the base measure $\mrd x$ as the
unknown and develop estimation and UQ methods based
on reweighting the sample values.  See \cite{tan:2004} for an
explanation.  The resulting methods are similar to 
frequentist methods that take $f$ as fixed and $\bsx_i$ as
random.  The next section compares the interval
estimates from MC, QMC and PN.

\section{Interval estimates}\label{sec:intervalest}
I consider the interval estimates from Monte Carlo,
based on the central limit theorem, to be `unreasonably
effective', despite some caveats in Section~\ref{sec:conc}.  
First, they are computable.  Second, they are even more
accurate than the estimate $\hat\mu$ is, so we actually
know more about our error than we do about the thing
we seek to estimate.

In plain MC with $w_i=1/n$, the error estimation is typically 
made based on the central limit theorem.
We can get statements like
\begin{align}\label{eq:mcuq}
\prt_\bsX\Bigl( |\hat\mu-\mu| > \frac{2.58 \hat\sigma}{\sqrt{n}}\mid f\Bigr) = 0.99 + O\Bigl(\frac1n\Bigr),
\end{align}
where $\hat\sigma$ is computed from $\bsX$.  The error term is $o(1/n)$ by the central
limit theorem but Edgeworth expansions in \cite{hall:1988} yield the given error term
assuming that $f(\bsx)$ has sufficiently many moments and is not supported on 
points of an arithmetic sequence.
Equation~\eqref{eq:mcuq}  shows that the error rate in the probability
statement is much better than the error rate in the estimate $\hat\mu$
itelf.  
If  we need more accuracy, perhaps because $n$ is small, 
the bootstrap-$t$ can get two-sided
interval estimates with error $O(1/n^2)$ \citep{hall:1988}
and that calibration is quite good even in
tiny samples \citep{elsmallsamp}.
Other bootstrap methods \citep{efro:tibs:1994} can get one-sided interval
estimates with error $O(1/n)$.

For QMC, the most studied 
counterpart to~\eqref{eq:mcuq} is the Koksma-Hlawka inequality
(see \cite{dick:pill:2010}) that gives
\begin{align}\label{eq:qmcuq}
 |\hat\mu-\mu| \le D_n^*(\bsx_1,\dots,\bsx_n)\times V_\hk(f)
\end{align}
where $V_\hk$ is the total variation of $f$ over $[0,1]^d$ in the sense of Hardy
and Krause. At first sight~\eqref{eq:qmcuq} looks like much better UQ than
\eqref{eq:mcuq} provides 
for MC.  There is no probability involved. Instead we get an absolute upper
bound on error and it holds for any integrand $f$ with $V_\hk(f)<\infty$.
Unfortunately the bound holding for all $f$ means it can be 
extremely conservative for some $f$.
Furthermore $D_n^*$ is extremely hard to compute
and $V_\hk(f)$ is much harder to get than $\mu$.  The upper bound
in~\eqref{eq:qmcuq} is then a product of two unknowns.
The comparison of~\eqref{eq:qmcuq} 
to~\eqref{eq:mcuq} calls to mind a point made by Ronald Fisher 
by way of George Barnard:
{\sl
In statistical inference, as distinct from mathematical inference, there is a world of difference between the two statements `p is true' and `p is known to be true'.
}

We can quantify uncertainty with~\eqref{eq:mcuq} but not with~\eqref{eq:qmcuq}.
Equation~\eqref{eq:qmcuq} remains valuable as
it shows that the MC rate can be improved 
via constructions achieving $D_n^*=O(n^{-1+\epsilon})$
for any $\epsilon>0$.

A counterpart to~\eqref{eq:mcuq} from Bayesian numerical analysis is
\begin{align}\label{eq:bnauq}
\prt_f\Bigl( |\hat\mu-\mu| > \frac{2.58 \hat\sigma}{\sqrt{n}}\mid f(\bsX)\Bigr) = 0.99,
\end{align}
where $\hat\mu$ and $\hat\sigma^2$ are the posterior mean and
variance of $\mu$ over randomness in $f$ given $f(\bsx_i)$.
This also looks better than~\eqref{eq:mcuq} because it has no
error term at all.  But we have reason to question whether
the probabilities in it are well calibrated.
The probability statement is ordinarily based
on a GP model. It is not an objective
Bayes statement because $f$ is not really sampled from the GP.
It is not quite a subjective statement either.  
The choice of GP usually takes into account qualitative properties of the GP 
such as mean squared differentiability that are 
satisfied by many different GPs that we could have chosen. 
From that set the selected GP is based largely on familiarity and 
computational feasibility, not just scientific opinion. 
Equation~\eqref{eq:bnauq} is not anybody's belief.


QMC accuracy can be combined with 
MC-based error quantification in randomized QMC (RQMC)
algorithms. One replicates an $n$ point QMC rule $m$ times.
RQMC is surveyed by  \cite{lecu:lemi:2002}.

Monte Carlo is unreasonably effective for error estimation
but in practice we use pseudo-random numbers.
That raises calibration issues due to flaws in the pseudo-random
number generators (PRNGs), which we turn to next.

\section{Testing and calibration}\label{sec:calib}

The numbers coming from a PRNG
are meant to simulate a stream of IID $\dustd(0,1)$ random
variables but they are not actually random.  That seems to 
place MC methods on the same footing as Bayesian
numerical analysis that treats a non-random $f$ as random.

Random number generators have been the subject
of thorough testing for several decades.  
New results still appear but the big crush in testU01 from \cite{lecu:sima:2007}
seems to have set the standard.
Some early PRNGs such as RANDU  \citep{lewi:good:mill:1969} 
had serious flaws but things are much better now. 
A flaw uncovered by \cite{ferr:land:wong:1992} was prominent
enough to make the news. The largest error in their
tables is $\hat\mu-\mu=0.000511$ where the
known value of $\mu$ was about $1.5$.
Pierre L'Ecuyer assures me (personal communication) 
that modern generators are better than the
ones used in that paper.
\cite{gelm:shir:2011} consider an average of $100$ independent 
draws from a posterior distribution, if we could get them,
to be sufficient in statistical applications because the 
numerical error comes along on top of a sizeable irreducible statistical error. 
\cite{vats:fleg:jone:2015} think larger samples are needed.
However, the point remains that errors from PRNGs not being
really independent uniform are not a serious problem
for those or most other MC applications.

By comparison, calibration for UQ modeling $f$ as random
is much less developed.  There is no `big crush' of problems
on which to calibrate Bayesian confidence interval methods (yet).
The calibration figures in this paper plot coverage probability versus
credibility level. It is encouraging that they show qualitative agreement that grows
better with increasing sample size.  
In applications we would like credible levels in the
half open interval $[0.99,1.0)$ and perhaps at $0.95$ as well.
The credible levels displayed
are $0$, $0.2$, $0.4$, $0.6$, $0.8$ and $1.0$.
Calibration at $100$\% should be automatically correct
so the most interesting results are at $80$\%
which is not high enough for cautious users.

The function $f$ is an infinite dimensional quantity,
and data may not `swamp the prior' in those
settings \citep{diac:free:1986}.
There are some signs that calibration will prove
hard for GP models in \cite{xu:stei:2017}. 
They consider functions $f(x)$ on $0\le x\le 1$.
If $f(x)=x^p$ is sampled at $x_i=i/n$ for $i=1,\dots,n$
and one uses a squared exponential covariance
model, then they conjecture that the maximum likelihood estimate of
the scale parameter is asymptotically proportional to $n^{p-1/2}$.
This holds theoretically for $p=0,1$ and it seems to hold empirically
for $p=0,1,2,3$ in their data.  A similar thing happened for
the easy case, but not the hard case,  in the authors' Figure 9.
We might have hoped for the
GP parameters to converge to some value, as it would if they
were being consistently estimated.  
Perhaps UQ calibrations can still converge properly in problems
where  a variance parameter converges to $0$ or diverges to $\infty$, but 
relying on that is worrisome.
Of course $x^p$ was
not drawn from the GP and getting that function has probability $0$.
On the other hand, any function that we might work
with has probability $0$ under the GP and we would want
calibration for it.

It is astonishing that PRNGs work as well as they do.
In practice floating point arithmetic not being the same as real
arithmetic causes more trouble.
We all owe a great debt to the people who did the
algebra and implementations behind modern PRNGs.




\section{Cubedness}\label{sec:cube}

The Bayesian approach to estimation and uncertainty quantification (UQ) 
typically includes a cost component proportional to $n^3$. 
That is a severe problem for integration methods.  If an integration method 
with error rate $O(n^{-\alpha})$ and cost $O(n)$
is to be replaced by a method with cost $O(n^3)$ the new 
method needs error rate $O(n^{-3\alpha})$ to be competitive (asymptotically).
If the method is far from competitive at estimating $\mu$
then its accuracy for UQ becomes much less well motivated. 
Users will ordinarily, though not universally, 
choose the method with greater accuracy 
over one with better UQ. 

To illustrate, suppose that plain MC can be run with some number $N=\eta n^{3}$
observations in the same time that PN with a GP can be run.  Then $n=\eta^{-1/3}N^{1/3}$. 
Let's use $\eta = 10^{-3}$. 
In Figure $6$ this value of $\eta$ would lead us to 
compare the QMC result with $N=2^{11}$ to the BC result with $n=127$ and 
we can substitute the one with $n=128$. If that is the right $\eta$, then plain QMC is much 
more accurate than BC.  Drawing Figure 6 with computational cost on the horizontal 
axis could leave it essentially unchanged or shift the QMC points to be above
$m/3$ or something in between.

For MCMC, suppose one uses $n$ observations 
and an $O(n^3)$ computing budget.  A competitor can 
run that MCMC for $2n$ observations, discard the first $n$
of them to get samples closer to the target distribution than the probabilistic numerics 
method would have.  Then the competitor can repeat that 
process independently some $O(n^2)$ times to greatly reduce the 
estimation variance by a factor like $O(n^2)$. Those replicates 
can also be used in UQ.   

There may be ways to mitigate the cubedness problem at 
least for  integration of smooth functions over $[0,1]^d$. 
\cite{jaga:hick:2018} reduce the cost to $O(n\log(n))$.
They do that by choosing $\bsx_i$ to be certain shifted lattice
points and then using also a special covariance kernel that
together with those input points allows fast transform methods to be used.

\section{Conclusions}\label{sec:conc}

\cite{henn:osbo:giro:2015} delivered a call to arms
for probabilistic numerical methods, as an alternative to classical methods.
The classical methods for integration are quite strong, making it a difficult
setting to score early improvements.
Those methods do however have weaknesses for integration,
and probabilistic methods could make a difference.
Some problems in engineering and climate modeling
have $f$ so expensive that the $O(n^3)$ cost
of algebra is much less than the cost of getting even one function
evaluation.  That removes most or all of the computational
advantage of classical methods.
Sometimes $f(\bsx)$ has an extremely skewed distribution
as for rare events, weakening the CLT,  and we cannot always find a good importance
sampler to compensate.  It can even happen that
$\int f(x)^2\rd x=\infty$ which makes the frequentist uncertainty
quantification problem extremely hard. 
\cite{peng:2004} has a good solution but even the best way to handle
heavy-tailed problems is not as good as having a light-tailed problem.
When the CLT is not available,
then much of the benefit of good random number generators
disappears with it.  With those three big advantages of the
classical method gone we might have to turn to the scienctific understanding behind the 
construction of $f$ to get a better answer. 
That puts the problem on grounds where Bayes has a
big advantage over classical alternatives. These
harder problems might not all be suitable for the plain Gaussian
process models that are central to probabilistic numerics at present.
That's a good thing because we need alternatives to those models
and new uses will appear for them once the alternatives develop.

I'll end with another reason for optimism about the probabilistic method.
\cite{sack:welc:mitc:wynn:1989}  find GP models to be more
accurate than response surface regressions.
In my experience, GP interpolation
has seemed unreasonably effective for approximation of functions such as those 
in the test bed of \cite{surj:bing:2014}. 
(I looked for survey articles to cite for this and saw
that not everybody had that same experience.)
When the GP approximation is working that well
and provides an easily integrable Bayesian approximation 
$\tilde f$ to $f$, we can write $f = \tilde f +(f-\tilde f)$
and integrate the two terms, using MC or RQMC
for the second term to get a better calibrated UQ.
This decomposition is a classical technique. \cite{ritt:2000}
gives it as Proposition II.4 and cites several earlier references.

\section*{Acknowledgements}

I thank Michael Stein,  Pierre L'Ecuyer, James Johndrow and Michael Lavine
for  answering some queries.  I thank the authors
for an interesting paper and the editorial staff of Statistical Science for 
the opportunity to comment.

\bibliographystyle{apalike}
\bibliography{comment}

\end{document}